\documentclass[galaxies,article,submit,oneauthor,pdftex,10pt,a4paper]{Definitions/mdpi} 
\usepackage{amssymb}
\firstpage{1} 
\pubvolume{xx}
\issuenum{1}
\articlenumber{5}
\pubyear{2018}
\copyrightyear{2018}
\history{Received: date; Accepted: date; Published: date}

\Title{A Multi-wavelength View of OJ 287 Activity in 2015-2017: Implications of
Spectral Changes on Central-engine Models and MeV-GeV Emission Mechanism}

\Author{Pankaj Kushwaha$^*$\orcidA{0000-0001-6890-2236}} 

\AuthorNames{Pankaj Kushwaha}
\address{%
\quad Department of Astronomy (IAG-USP), University of Sao Paulo, Sao Paulo 05508-090, Brazil\\
\quad Aryabhatta Research Institute of Observational Sciences (ARIES), Nainital 263002, India}

\corres{Correspondence: pankaj.tifr@gmail.com}

\abstract{A diverse range of observational results and peculiar properties across
the domains of observation have made OJ 287 one of the best-explored BL Lac objects
on the issues of relativistic jets and accretion physics as well as the strong
theory of gravity. We here present a brief compilation of observational results from the
literature and inferences/insights from the extensive studies but focus on the
interpretation of its $\sim$ 12-yr quasi-periodic optical outbursts (QPOOs) and
high energy emission mechanisms. The QPOOs in one model are attributed to the 
disk-impact related to dynamics of the binary SMBHs while alternative models attribute
it to the geometrical effect related to the precession of a single jet or double jets.
We discuss implications of the new spectral features reported during the 2015--2017
multi-wavelength high activity of the source -- a break in the NIR-optical spectrum and
hardening of the MeV-GeV emission accompanied by a shift in the location of its peak,
in the context of the two. The reported NIR-optical break nicely fits the description
of a standard accretion disk emission from an SMBH of mass $\sim~10^{10}~M_\odot$ while
the time of its first appearance in end-May 2013 (MJD 56439) is in close coincidence with the
time of impact predicted by the disk-impact binary SMBH model. This spectral and
temporal coincidence with the model parameters
of the disk-impact binary SMBH model provides independent evidence in favor of the
model over the geometrical models which argue a total central-engine mass in the range
of $\rm 10^{7-9}~M_\odot$. On the other hand, the MeV-GeV spectral change is naturally
reproduced by the inverse Compton scattering of photons from the broad-line region and
is consistent with the detection of broad emission lines during the previous cycles
of quasi-periodic outbursts. Combining this with previous SED studies suggests that in
OJ 287, MeV-GeV emission results from external Comptonization.}

\keyword{BL Lac objects: individual: OJ 287 -- galaxies: active -- galaxies: jets -- radiation
mechanisms: non-thermal-- gamma-rays: galaxies}

\begin{document}

\section{Introduction}
Non-thermal emission is ubiquitous with jets hosted by compact astrophysical objects
being one of the most prominent emitters. One of the most powerful and persistent
jets are those found in active galactic nuclei (AGNs) designated radio-loud, characterized
by a large scale, highly collimated, relativistic jets of plasma. In particular,
the subclass called blazar -- AGNs with bipolar relativistic jets of plasma aligned
to our line of sight has its entire emission almost fully dominated by the jet emission.
They have been found to emit across the entire electromagnetic (EM) spectrum, from radio
to GeV-TeV energies \citep[$\sim$ 17--20 orders of magnitude; e.g.][]{2010ApJ...716...30A},
variable on all timescales from orders of a few 10s of minutes and even less
to decades\footnote{currently feasible with observing facilities and available data}
\citep[$\sim$ 6--7 orders of magnitude; e.g.][]{2018ApJ...863..175G}. Their radio
and optical emission is highly polarized and have been observed to vary often with
the source flux, all the way from 0 to $\rm > 50$\%. Imaging in radio, infra-red (IR),
optical, and X-ray, on the other hand, show an extremely well collimated jet extending
up to Mpc scales \citep[$\sim$ 9--13 orders of magnitude; e.g.][]{2006ApJ...648..910U,
2011ApJ...729...26M} with frequent sighting of superluminal features in high-resolution
imaging of the core. Taken
together with the observational and theoretical understanding of other astrophysical
objects combined with their high and rarely repeating observational behavior indicates
that they are the site of complex, multi-level, multi-scale physics and thus, are normally
called extreme sources among non-catastrophic events.

Though studies in different energy bands finally culminated into a unified scheme
for radio-loud AGNs \citep{1995PASP..107..803U}, it also revealed that AGNs need
well-coordinated multi-wavelength (MW) observations to go beyond the limits
associated with individual bands. The launch of the gamma-ray survey observatory
\emph{Fermi}-LAT (Large Area Telescope) with an unprecedented MeV-GeV sensitivity
\citep{2009ApJ...697.1071A} has revolutionized studies of \(\gamma\)-ray
emitting sources, especially blazars. The \emph{Fermi} AGN monitoring program\footnote{Multi-wavelength support program -- \url{https://fermi.gsfc.nasa.gov/ssc/observations/multi/programs.html}} supported by a host of observatories across the globe, working in
different energy bands and supplementing and coordinating \emph{Fermi} observations
currently serves as the best archive of (relatively) unbiased data. The \emph{Fermi}-LAT
survey catalog, as was expected,
revealed blazars (and AGNs) as the largest source population, making $>$60\% of
the detected sources \citep[4FGL;][]{2019arXiv190210045T}. This allowed
a detailed spectral, temporal, polarization, and imaging characterization of blazars,
as well as exploration of correlations between them, thereby greatly enhancing our understanding of these sources and physical conditions within the relativistic jets.

Spectral and temporal studies, in particular, high energy (HE) emission mechanisms
and search for (quasi) periodic modulation has been one of the focus of intensive
research, in addition to characterization based on these. Temporal studies of flux
variability have firmly established it to be stochastic with statistical properties
\citep{2018ApJ...863..175G,2017ApJ...849..138K,2016ApJ...822L..13K} broadly similar
to those exhibited by other-accretion powered sources \citep{2015SciA....1E0686S}.
Though a few cases of firm quasi-periodic signals have been observed, significance
of most of the detected signal is still marginal \citep[e.g.][and references therein]{2019MNRAS.484.5785G}.
Broadband spectral studies, on the other hand, revealed a characteristic broad
double-humped spectral energy distribution (SED). This feature culminated in a 
new classification scheme for
blazars based on the frequency ($\nu_p$) at which the low-energy component peaks \citep{1998MNRAS.299..433F,2010ApJ...716...30A}. Thus,
sources with $\nu_p < 10^{14}$ Hz, $10^{14}< \nu_p < 10^{15}$, and $\nu_p > 10^{15}$
are respectively called as low-synchrotron peaked (LSP), intermediate-synchrotron
peaked (ISP), and high-synchrotron peaked (HSP) blazars. So far only {BL Lacs
(BLLs) subclass of blazars} have been found to exhibit the three spectral classes,
referred to respectively as low-frequency-peaked BL Lacs (LBLs), intermediate-energy-peaked
BL Lacs (IBLs), and high-frequency-peaked BLLs (HBLs) \citep{1998MNRAS.299..433F} whereas
 flat spectrum radio quasars (FSRQs) so far are exclusively LSP sources.

The low-energy hump of the blazars SED extends from radio to ultraviolet(UV)/X-ray
energies, attaining a maximum in between NIR to soft-X-rays and is widely regarded
as the synchrotron emission from a relativistic non-thermal electron in the jet. The
high-energy hump spans X-rays to GeV--TeV energies, peaking in MeV-GeV energies,
but its origin remains uncertain. The uncertainty is a direct reflection of lack
of constraint on the matter content of the jet plasma, whether mainly electrons
(leptonic) and/or hadrons (protons primarily) and particle acceleration. Within the
limit of current observational constraints, both the models have been successful
in explaining observed broadband SEDs \citep[e.g.][]{2012rjag.book.....B,2013ApJ...768...54B,
2015MNRAS.450L..21Z,2008MNRAS.387.1669G} though the exact cause and the level of 
contribution/dominance remains a matter of debate. In the leptonic scenario, entire
broadband emission is attributed to the primary electrons, via synchrotron at radio to
UV/X-ray energies and via inverse Compton (IC) scattering from X-ray up to TeVs
\citep[e.g.][]{2008MNRAS.387.1669G,2014Natur.515..376G,2007ApJ...669..862A}. Until
recently, this has been the most favored interpretation
of blazar emission due to the photon rich environment offered by the AGNs constituents
for IC scattering e.g. accretion disk (AD) close to the source, broad-line region
(BLR) with extension up to sub-parsec scales, infra-red torus on parsec (pc) scales,
and the omnipresent cosmic microwave background (CMB), in addition to the synchrotron
photons. The respective IC scattered radiation, in blazar community, are referred to
as EC-AD \citep{1992A&A...256L..27D}, EC-BLR \citep[][]{1994ApJ...421..153S},
EC-IR \citep{1994ApJ...421..153S}, EC-CMB, and synchrotron self-Compton \citep[SSC;][]{1996ApJ...461..657B} where EC stands for External Comptonization -- IC scattering
of photon field external to the jet. Depending on the location of the emission region,
one or many of the photon fields can contribute and/or dominate the scattering process.
In general, FSRQs requires EC to explain their \(\gamma\)-ray emission while SSC
is sufficient for most of the BLLs (IBLs and HBLs). Given the current understanding of emission
lines strength in the sub-classes of blazar \citep[and references therein]{2012ApJ...751..159A}, it seems that to a broad basic level, the traditional sub-classification of blazars
based on rest-frame equivalent width (EW) of optical emission lines into BL Lacartae
objects (BLLs, EW $< 5$ \AA{}) and flat spectrum radio quasars (FSRQs, EW $> 5$
\AA{}) nicely integrate into the interpretation that in BLLs the high-energy hump
is primarily powered by SSC while its by EC in FSRQs. 

The hadronic scenario attributes the HE hump to proton synchrotron (purely electromagnetic
process) and/or cascade initiated as a result of interaction of ultra-relativistic 
protons ($\geq EeV$) with photons and/or protons\footnote{being
the biggest constituent, almost exclusively} \citep[e.g.][]{2013ApJ...768...54B}.
The recent detection of a $\sim$ 290-TeV neutrino
(IceCube-170922A) by the \emph{IceCube} observatory in the direction of blazar TXS
0506+056 \citep{2018Sci...361..147I,2018Sci...361.1378I}, spatially and temporally
coincident with its flaring MW activity provided the first clear evidence in favor of hadronic emission. Further investigation of data revealed neutrinos detection in the
same direction during a quiescent state of the source \citep{2018Sci...361..147I}.
Though interpretations of these neutrinos are still under debate \citep[and references
therein]{2019arXiv190712506M}, modeling of SEDs corresponding to the neutrino episodes suggests an
overall sub-dominant hadronic contribution \citep{2018NatAs.tmp..154G}.

Majority of the blazars MW activities have been observed to be simultaneous within
the observational cadence. With radio to optical emission being synchrotron,
the standard leptonic scenario provides the simplest, natural and logical explanation
to such highly correlated variability though more complex physical processes may
be involved and thus, more involved interpretations can be offered \citep[e.g.][]
{2018MNRAS.479.1672K}. For BLLs, in the standard leptonic picture, an SSC interpretation
has been generally favored in the literature for their HE hump due to a weak or
absent BLR field. However, increasing number of studies suggest that this may not
be true for LBLs and instead argue EC-IR for the MeV--GeV emission
\citep{2013MNRAS.433.2380K,2012ApJ...751..159A,
2018A&A...616A..63A}. For example, in the case of 2009 flare of LBL OJ 287,
\citet{2013MNRAS.433.2380K} have shown the in-feasibility of SSC in explaining
MeV--GeV emission through a systematic modeling of SEDs within the observational
constraints while for LBL AO 0235+164, \citet{2012ApJ...751..159A} have argued the
EC interpretation based on energetics and the luminosity of the detected emission
lines. The inability of SSC to reproduce the X-ray and $\gamma$-ray emission in the
case of OJ 287 was already apparent from the work of \citet{2009PASJ...61.1011S}, with
the only exception being the lack of a contemporaneous MeV-GeV spectrum. The EC-IR
interpretation of MeV-GeV gamma-ray emission in these two is consistent with
inferences drawn from the study of broadband SEDs of a sample of LBLs with good
quality $\gamma$-ray spectra from \emph{Fermi}-LAT \citep{2018A&A...616A..63A}.
A similar (EC-IR) interpretation is favored for FSRQs from the lack of spectral
cutoff at $\gtrsim$ 20 GeV, expected due to the $\gamma-\gamma$ interaction,
in the spectra extracted from $\sim$ 7.3-yr of \emph{Fermi}-LAT \citep{2018MNRAS.477.4749C}.
However, this interpretation has an important caveat in the case of FSRQs. FSRQs
are believed to have a rich IR torus field and the expected spectral cutoff due to
$\gamma-\gamma$ interaction for it will occur at very high energies (VHEs, E $>$
100 GeVs). Thus, spectra extracted from data integrated over a long duration may
contain moments when the emission has mainly happened at parsec scales (EC-IR) for
which the mentioned cutoff will not lie in \emph{Fermi}-LAT band and thus, may
hide/suppress the cutoff feature. 

In the present work, we focus on the blazar OJ 287 -- a potential binary SMBH
candidate system. In particular, we discuss the implications of
the new spectral features observed during its 2015-2017 MW activity on the source
central engine models discussed in the literature and high energy emission (MeV --
GeV) mechanisms -- one of the fundamental debate in the blazar community and focus
of intensive studies. First, we present a brief historical account of the source
general properties as gleaned from observations and models/interpretations
offered, if any, for its unique features in the next section. In \S\ref{sec:discussion},
we discuss the implications of the reported features in the context of the main theme
of the work with summary in \S\ref{sec:summary}. We have assumed a $\Lambda$CDM
cosmology with $\Omega_M = 0.286$, $\Omega_\Lambda = 0.714$, and a Hubble constant
$\rm H_0 = 69.6$ km/s/Mpc. With this,  OJ 287 redshift of 0.306 corresponds to
a luminosity distance of 1.6 Gpc and an angular diameter scale of 4.556 kpc/arcsec.

\section{OJ 287}\label{sec:oj287}
OJ 287, as it is called now, was first reported in the second section of the VRO
(Vermilion River Observatory) survey at 610.5 MHz \citep[VRO 20.08.01;][]
{1967AJ.....72..757D} and its optical association was identified by 
\citet{1970ApL.....6..201B}. Spectroscopic attempts following its identification failed to
reveal any emission and or absorption features \citep{1971ApL.....9..147K,
1973ApJ...185..145V} and was firmly established only later at z = 0.306 when it
was in a low flux state \citep{1985PASP...97.1158S,1989A&AS...80..103S}. However, photo-polarimetric studies
in radio to optical bands, in general, found it to be similar to other radio sources
with a star-like appearance in optical images, inverted radio spectrum, variable
brightness, polarized and variable non-thermal continuum \citep{1971ApL.....9..147K,
1971ApL.....9..151A,1971NPhS..234...71D,1972ApJ...178L..51E,1973ApJ...179..721V,1973ApJ...185..145V}. These findings and resemblance of its flux and polarization variations with the archetypal source BL Lacartae, it was classified as a BL Lac
type object \citep{1971ApL.....9..147K}. These observations also found it to be
the most dynamic among all, with comparable variations over a broad range of wavelengths
\citep[radio and optical; ][]{1971ApL.....9..147K}. As a result, OJ 287 became the key
source to characterize and understand the BL Lac class \citep{1985PASP...97.1158S}. 
Subsequently, many focused as well as coordinated multi-wavelength studies were
carried out in both photometric and photopolarimetric modes
\citep[e.g.][]{1971NPhS..234...71D,1984MNRAS.211..497H,1973ApJ...179..721V,1973ApJ...185..145V}.

Initial concerted efforts revealed a tantalizing $\sim$ 39.2-minute periodic signal
in the optical observation while historical data archive revealed data going back to
1890 with four clear high activity duration of several months \citep{1973ApJ...179..721V}. OJ 287, however, became famous after the discovery of persistent quasi-periodic
outbursts (QPOs) of $\sim$ 12-yr in the optical band by \citet{1988ApJ...325..628S}. Various
models have been proposed in the literature for this repeating modulation and can be
broadly grouped into two classes: dynamical and geometrical.
The dynamical models attribute the recurring outbursts to accretion dynamics
in a supermassive binary black holes (SBBHs) system \citep{1988ApJ...325..628S,
1996ApJ...460..207L,2008Natur.452..851V} while geometrical models attribute it
to Doppler boosted jet emission resulting from the jet precession \citep[e.g.][and
references therein]{1997ApJ...478..527K,2018MNRAS.478.3199B,2018arXiv181111514Q}.

The very first model by \citet{1988ApJ...325..628S} postulated OJ 287 as an SBBH
system. The outbursts in the model were result of enhanced accretion to the primary as a result of perturbation 
caused by the secondary when transiting the periastron in an orbit coplanar with 
the primary's accretion disk. It was based on the similarity of flare profiles and
its structure to that of an accreting system \citep{1985A&A...147...67S}. The model successfully passed its first observational test by correctly predicting the 1994
optical outburst \citep{1996A&A...305L..17S}. However, the intensive monitoring
of the 1994 outburst revealed double-peaked flares with rather sharp substructures
within it compared what was expected from an accretion flow \citep{1996A&A...315L..13S}.
Further, \citet{1988ApJ...325..628S} model failed to explain the twin
nature of the outbursts. This led \citet{1996ApJ...460..207L} to propose
the twin outbursts to be due to the impact of the secondary SMBH on the accretion
disk of the primary twice every orbit (see also \citet{1997ApJ...484..180S}). It
attributed the outbursts to hot bubble of gas, torn off from the disk as a result of the impact
which when expands and become optically thin, emits strongly in optical-UV
bands via thermal bremsstrahlung. The flare emission is thus
completely unpolarized. Another model in the dynamical class is by \citet{2013MNRAS.434.2275T}.
It is based on theoretical studies and numerical hydrodynamic/magneto-hydrodynamic
simulations of nearly equal-mass binary SMBH systems going around each other in an
orbit coplanar with their circumbinary disk. For OJ 287, it assumes a total mass of $\sim~10^9~
M_\odot$. The QPOs occur due to leakage of gas into this cavity
once every orbit. Though the model generates
double flares, it has not been investigated in detail vis-a-vis the vast
amount of available observational data on OJ 287. Thus, though it remains a plausible
interpretation, the details are still to be worked out and tested with the
available data.

Several other models, mainly based on geometrical interpretations have been proposed
in the literature with earliest ones based solely on the recurrent double-peaked
optical outbursts \citep{1997ApJ...478..527K,1998MNRAS.293L..13V} while subsequent
ones considering other observed inputs of OJ 287 like radio measurements
\citep{2000ApJ...531..744V,2007Ap&SS.310...59S}, optical polarization 
\citep{2010MNRAS.402.2087V}, morphological changes \citep{2018MNRAS.478.3199B} and
jet kinematic features on parsec scales \citep{2018arXiv181111514Q,2015RAA....15..687Q}.
Except for \citet{2000ApJ...531..744V} and \citet{2010MNRAS.402.2087V} models, the recurring
outbursts in these models are attributed to the Doppler boosting caused by systematic changes in the jet 
orientation with respect to our line of sight as a result of precession.
In the model of \citet{1997ApJ...478..527K}, the gravitational torque of the secondary
induces precession in the primary's accretion disk which in turn leads to jet precession
and cause the first flare. It attributes the second outburst to the nodding motion.
\citet{1998MNRAS.293L..13V} model argue that both the SMBH
have precessing relativistic jets. The interaction between jet plasma and the ambient
medium leads to the bending of the jets and the double-peaked outbursts arise when the
bent jets are aligned towards us as a result of precession.
\citet{2000ApJ...531..744V} proposed a hybrid interpretation for the twin outbursts
based on the radio variability where they found that the first optical outburst has no
radio counterpart while the second shows simultaneous radio and optical flares.
They attributed the first flare to the disk impact and the second to the jet, resulting
from the propagation of impact disturbances to the primary's jet. Another
double jet model is proposed by \citet{2007Ap&SS.310...59S} by additionally
considering variability in radio and ``double minimum'' in the optical data. They
interpret the double-peaked flares to the synchrotron emission from the double
helix jet which appears partially merged at radio, giving rise to a broad temporal
profile in radio bands.

The most recent and detailed models in the series of geometrical interpretations
are by \citet{2018MNRAS.478.3199B} and \citet{2018arXiv181111514Q}.
\citet{2018MNRAS.478.3199B} found a $\sim$ 22-yr periodicity in the morphological
features of the parsec-scale jet. Assuming no abrupt changes in these kinematic features,
they interpreted both the recurrent outbursts in optical and radio bands to Doppler
boosted jet emission as a result of precessing and jet rotation. They further argue
that a binary system is not needed and even a precessing accretion disk can generate
the jet precession. \citet{2018arXiv181111514Q}, on the other hand, showed that the 
trajectories of superluminal radio knots seen at 15 GHz and 43 GHz can be explained
in a binary SMBH scenario with both SMBHs having precessing jets. Using the correlation
between OJ 287 high activity and ejection of superluminal knots, \citet{2018arXiv181111514Q} 
further showed that the QPOs can be explained by the motion of a
few of these knots. A completely different interpretation for the recurrent outbursts
is proposed by \citet{2010MNRAS.402.2087V}
from the study of optical photopolarimetric properties from 2005 to 2009. They
attributed the observed features to the 'magnetic breathing' of the disk causing
accretion of magnetic field lines and claimed that there should be no more such
outbursts in the future.

As stated above, OJ 287 is a very dynamic source and the diversity of peculiar
features have made it the
best-monitored BLL blazar over a wide range of scales in all the domains of observation
e.g. spectral: \citep{1971NPhS..234...71D,2000A&AS..146..141P,2018MNRAS.479.1672K,
2018MNRAS.473.1145K,2013MNRAS.433.2380K}; temporal: \citep{1988ApJ...325..628S,
2000A&AS..146..141P,2013A&A...559A..20H,2016ApJ...819L..37V,2017MNRAS.468..426S,
2018MNRAS.480..407K,2018ApJ...863..175G}, Imaging: \citep{2011ApJ...729...26M,
2012ApJ...747...63A}; polarization:\citep{2019AJ....157...95G,2018ApJ...862....1C,
2009ApJ...697..985D,2000A&AS..146..141P} on the aspects of jet, accretion physics,
and test of the general theory of relativity. Below we list different
observational facets of the source from the literature in each of the domains and
inferences from these. It should,
however, be noted that for some of the domains there is no clear boundary due to
the method of detection used in some of the energy bands (e.g. radio measurement are
primarily imaging and thus observation at different epoch provide temporally-sequenced
data, thereby mixing the two). Further, the list of literature cited/mentioned in 
the context of observational features is neither exhaustive nor complete, but
rather a practical consideration. In addition to citing studies reporting new
features/properties and/or performing extensive analysis, we have followed a simple guiding
principle where only references with earliest and latest observations are mentioned
if the reported features/properties are similar.

\subsection{Spectral}
The general broadband SEDs of the source is typical of LSP/LBL subclass of
blazars with emission up to GeV energies \citep{2010ApJ...716...30A,2013MNRAS.433.2380K}.
Reported SEDs suggest the low-energy component peak at $\lesssim$ NIR bands and
$\lesssim$ 100 MeV for the high-energy component \citep{2000A&AS..146..141P,
2010ApJ...716...30A,2013MNRAS.433.2380K}. So far no shift in the location of the
peak of the low-energy component has been observed, though additional emission
component, similar to an HBL SED \citep{2018MNRAS.479.1672K} was recently seen
during its first-ever reported VHE activity \citep{2017arXiv170802160O}. Studies
in different energy bands at different flux states of the source, however, have
reported a wide variety of spectral features.

\begin{figure}[!ht]
\includegraphics[scale=1.2]{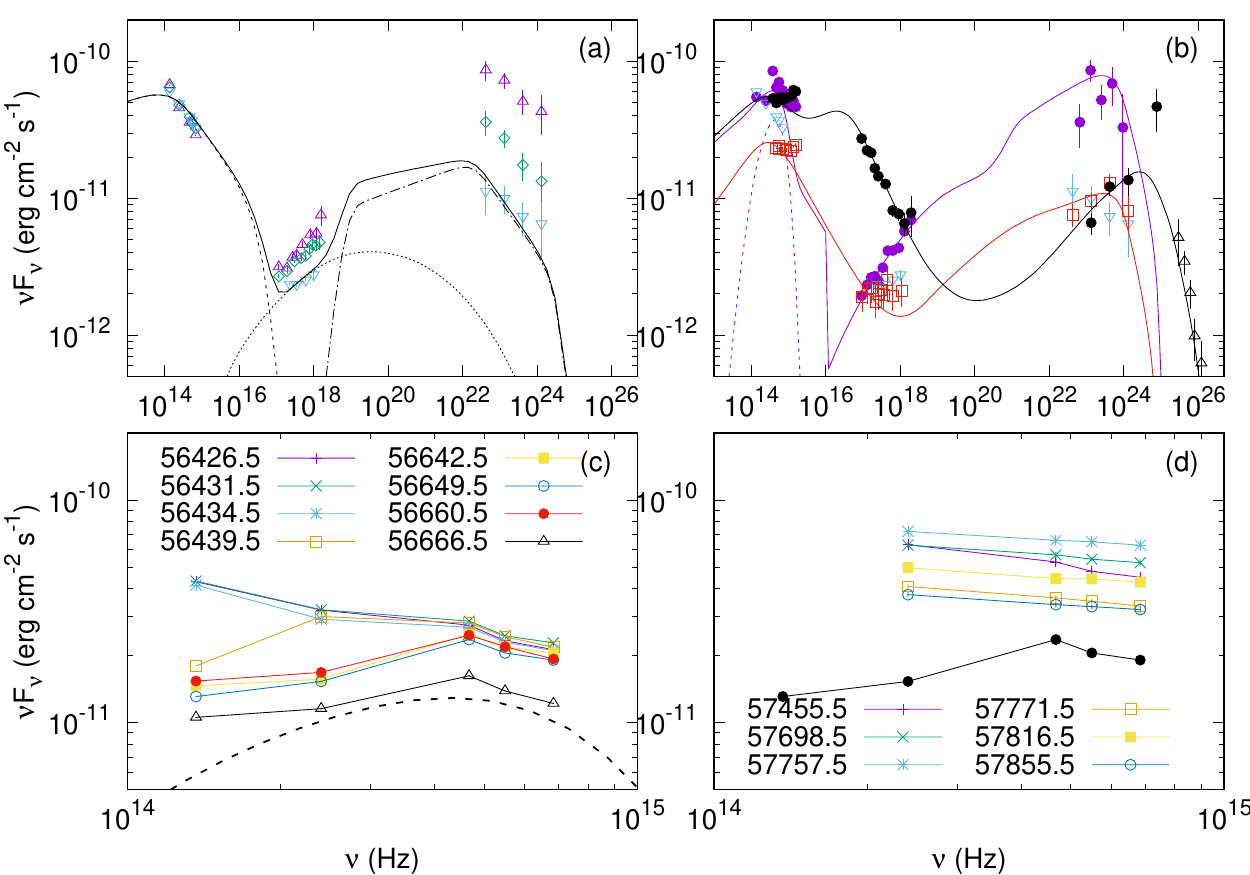}
\centering
\caption{Broadband and optical SEDs showing the spectral phases exhibited by OJ 287
to date.  (\textbf{a}) Typical broadband SEDs of OJ 287 from a 2009 observation
\citep{2013MNRAS.433.2380K} showing three different flux states: Flare (magenta,
MJD: 55124-55131), moderate (green, MJD: 55131-55152) and quiescent (cyan, MJD:
55152-55184) in MW emission. The solid curve is the total emission with synchrotron,
SSC and EC-IR component shown respectively in dashed, dotted, and dot-dashed curves
\citep[see][for modeling details]{2013MNRAS.433.2380K}. (\textbf{b}) Broadband SEDs
of the source in its new spectral phase during 2016-2017 with magenta (MJD: 57359-57363;
\citep{2018MNRAS.473.1145K}) showing a flare SED, black (MJD: 57786; 
\citep{2018MNRAS.479.1672K}) showing a typical SED during the VHE phase and red
(MJD: 57871; \citep{2018MNRAS.479.1672K}) showing the quiescent SED state after
the VHE activity with the lowest flux across EM spectrum. For reference/comparison,
the quiescent SED (cyan) from panel (\textbf{a}) is also shown. The solid curves
are the model produced spectrum \citep[for details]{2018MNRAS.479.1672K,2018MNRAS.473.1145K}
while the dotted curve is the standard accretion-disk spectrum of a \(\sim 
10^{10} ~M_\odot\) SMBH. (\textbf{c}) NIR-optical spectrum highlighting the timing
of tbe appearance of spectral break \citep{2018MNRAS.473.1145K}. The dashed curve is the
accretion-disk spectrum drawn again for clarity. (\textbf{d}) NIR-optical SEDs
before (MJD 57455.5), during and after the VHE activity of OJ 287, showing return to
the typical power-law NIR-optical spectrum. For comparison, one of the SED (black)
from (\textbf{c}) is also shown.}
\label{fig:msSEDs}
\end{figure}

At radio centimeter wavelengths, it shows a convex (positive curvature) spectra
\citep{1970ApL.....6..201B}, typical of the synchrotron self-absorbed quasi-stellar
sources. This changes to power-law spectra representing optically thin synchrotron
emission at mm wavelengths \citep{1973ApJ...185..145V,
1982ApJ...261..403W}. At NIR-Optical-UV energies, reported spectra can be separated
into two groups. The first group corresponds to the duration before mid-2013
(MJD $<$ 56439) when data in these bands are smoothly connected, irrespective
of the source flux state and occasionally show hints of smooth curvature at either
or both ends \citep{1971NPhS..234...71D,1973ApJ...185..145V,1982ApJ...261..403W,
1986Natur.324..546G,1998A&AS..133..353H, 2000A&AS..146..141P,2009PASJ...61.1011S,
2013MNRAS.433.2380K,2017MNRAS.468..426S,2019AJ....157...95G}. This includes spectra
measured during the epochs of outbursts claimed to have a thermal origin in the
binary SMBH model of
\citet{1996ApJ...460..207L}. All these are consistent with synchrotron emission
from a power-law particle distribution. From mid-2013 (May 2013, MJD 56439, fig. \ref{fig:msSEDs}-c) till February
2016 (MJD $\sim$ 57455; fig. \ref{fig:msSEDs}-d), a spectral break was observed at the junction of the NIR-optical
region \citep{2018MNRAS.473.1145K}. After February 2016, it again returned to its
typical pre mid-2013 spectral state.

In contrast, at X-ray energies, it has been observed with most of the major X-ray
facilities and has exhibited drastic spectral variations compared to
the other EM bands. The observed spectra cover all the possible spectral phases.
In addition to its typical LBL spectra described by a power-law photon spectral
index of $\Gamma\sim$
1.5-1.7 \citep[$f_\nu \sim \nu^{-\Gamma}$;] []{1997PASJ...49..631I,2001PASJ...53...79I,
2009PASJ...61.1011S,2013MNRAS.433.2380K,2018MNRAS.473.1145K}, it has shown a flat
\citep[$\Gamma \sim ~1$;][]{2017MNRAS.468..426S,2013MNRAS.433.2380K,
2018MNRAS.479.1672K,1996A&AS..120C.599S}, extremely soft $\Gamma > 2$ \citep[]
[and references therein]{2017arXiv170802160O,2018MNRAS.479.1672K,1997PASJ...49..631I}
as well as mixture of these \citep{2001PASJ...53...79I,2018MNRAS.479.1672K,
2019arXiv191202730P}.
Interestingly, most of the extremely soft spectral state seems to have been around (within
a few years) the period of optical outbursts claimed to be thermal bremsstrahlung
emission \citep{1996ApJ...460..207L,2018ApJ...866...11D}. The latest 
steep X-ray spectral state was seen during the VHE activity of the source in
2017 \citep{2017arXiv170802160O} which also corresponds to the highest-ever
reported X-ray emission from OJ 287 \citep[and references therein]{2018MNRAS.479.1672K}.

Spectral inferences at gamma-rays before \emph{Fermi}-LAT have only been indirect.
A probable detection was claimed in the \emph{Compton Gamma-Ray Observatory} (CGRO)
energy band on board the \emph{EGRET} during a high optical state of the source
in 1994 \citep{1996A&AS..120C.599S}. Though photon statistics prevented spectral
analysis, it was argued to be in a ``hard'' state based on the number of detected
GeV photons. In the \emph{Fermi}-LAT era, on the contrary, it is one of the bright
MeV-GeV sources detected in the first 3-month of its operation with spectrum being
consistent with a power-law profile \citep{2010ApJ...716...30A}. Till date, all the
reported MeV-GeV spectra before its latest activity starting November 2015 show a
power-law \citep{2013MNRAS.433.2380K} profile and suggest SED peak at $\lesssim$
100 MeV. The November 2015 MW activity revealed a hardened MeV-GeV spectrum with a
shift in the HE SED peak and yet to revert to its generic form as per the latest
records \citep{2017arXiv170802160O,2018MNRAS.473.1145K,2018MNRAS.479.1672K}. It
also registered its first VHE activity \citep{2017arXiv170802160O} in 2016, accompanied
by a change in MeV-GeV spectral state which is consistent with the extrapolated
VHE spectrum \citep{2018MNRAS.479.1672K}. The VHE activity was a transient phase,
lasted about six months with source HE spectrum being of an LBL+HBL source
\citep{2018MNRAS.479.1672K}. Following this, OJ 287 has been added to the TeV source
catalog\footnote{http://tevcat.uchicago.edu/}. A snippet of different spectral states
exhibited by OJ 287 is shown in Figure \ref{fig:msSEDs}.

\subsection{Temporal}
OJ 287 has shown variability on all time scales (e.g. minutes-to-hours:
\citep{2000A&AS..146..141P,2017MNRAS.465.4423G}, days: \citep{2000A&AS..146..141P,
2013MNRAS.433.2380K}, months-to-years: \citep{1973ApJ...179..721V,2000A&AS..146..141P,2016ApJ...819L..37V},
decades: \citep{1988ApJ...325..628S,2013A&A...559A..20H}) across the entire
electromagnetic spectrum \citep[e.g.][]{1982ApJ...261..403W,2013MNRAS.433.2380K,
2018MNRAS.473.1145K,2018MNRAS.479.1672K}. On short time scales ($<$ months), it
normally exhibits variation from NIR to MeV-GeV $\gamma$-ray energies which have been
simultaneous within the observational cadence
\citep[e.g.][]{2018MNRAS.479.1672K,2018MNRAS.473.1145K,2013MNRAS.433.2380K}.
An exception to this was the August 2016 -- July 2017 (MJD: 57600 -- 57950) period
when simultaneous variation was seen only at optical to X-rays energies but 
statistically no variability at Fermi-LAT energies \citep{2018MNRAS.479.1672K}. 

On long terms, MW variations, in general, are more pronounced
at optical energies than X-rays but show no relation/pattern between the two
\citep{2017MNRAS.468..426S}. 
In the Fourier domain, MW variations show flicker/colored-noise spectra on all
timescales with $\gamma$-ray variation being different from the others. 
Statistical analyses of the radio to X-ray light curves show colored-noise
power spectral densities (PSDs) for all and consistent with each other. Gamma-ray
time series from LAT, on the other hand, shows both white and
color noise PSD with the transition occurring on a timescale of 150-days
\citep{2018ApJ...863..175G}.  Apart from stochastic variations, OJ 287
has also exhibited (quasi) periodic
modulation and in fact, is the only blazar/AGN with the most numerous claims of QPOs
of different duration in different energy bands e.g. optical: \(\sim 40\) minutes
\citep{1973ApJ...179..721V,1974MNRAS.168..417F}, 0-50 days \citep{2013MNRAS.434.3122P}, $\sim$
400-days \citep{2016ApJ...832...47B}, $\sim$ 11.65 years \citep{1988ApJ...325..628S,
2018MNRAS.478.3199B}; radio: \(\sim 16\) minutes \citep{1985Natur.314..148V},
1.12 up 6-7 years \citep[and references therein]{2018MNRAS.478.3199B}.

\subsection{Imaging}
At radio wavelengths, OJ 287 exhibits a one-sided jet with complex patterns on
parsec \citep{1987ApJ...323..536R,2012ApJ...747...63A} and kilo-parsec scales.
Morphologically, it appears as an FR I source but energetically exhibits an FR II
power. On Very Long Baseline Array (VLBA) resolution scales, it presents very complex
dynamical patterns \citep{2012ApJ...747...63A} with a core-jet structure
\citep{1987ApJ...323..536R,2004ApJ...608..149T}, changes in jet position angle as
well as overall morphology on parsec scales \citep{2004ApJ...608..149T,2012ApJ...747...63A,
2018ApJ...862....1C}. \citet{2004ApJ...608..149T} study of kinematic features at
8 GHz has reported a clockwise change of $\sim ~30^\circ$ in jet position angle
and argue for a ballistic precessing jet for this while \citet{2012ApJ...747...63A}
have reported a sharp swing in jet position angle by $> 100^\circ$ within the 0.4mas
($\sim$ 1.8 pc) inner region between 2004-2006 at 43 GHz. Based on structural changes 
\citet{2012ApJ...747...63A} have instead argued for an erratic wobbling of the jet along with superluminal
motions in non-radial directions. \citet{2017Galax...5...12C} and \citet{2018MNRAS.478.3199B}
study of VLBA observations at 15 GHz, on the contrary, argue for a precessing jet
with a period of $\sim$ 30-yr and $\sim$ 22-yr respectively. From the analysis of
the ridgeline of contours, the former has argued the jet to be a rotating helix.
In contrast, a reanalysis of the same by data \citet{2018MNRAS.478.3199B} in combination
with other observations have reported additional yearly variations under the assumption
that observed changes are smooth. Based on these morphological changes, \citet{2018MNRAS.478.3199B} have
favored a precessing jet for the cause and claims that this can explain both radio
and optical variations within this framework. Correlation studies between mm radio
time series with other bands (optical to $\gamma$-rays) have found a strong connection
between these during the high activity periods with the increased activity associated with
radio-flaring in quasi-stationary features and ejection of superluminal features
\citep{2011ApJ...726L..13A,2017A&A...597A..80H}.

At X-ray energies, deep exposure with the \emph{Chandra} observatory shows a curved
jet with a de-projected extension of $>$ 1 Mpc and many bright X-ray knots/hot-spots.
The location of these knots are consistent with the radio ones but follow a complex
brightness profile while only the bright core is visible at NIR-optical energies
\citep{2011ApJ...729...26M}. Claims of detection of the host galaxy, so far, is 
been ambiguous \citep{1998MNRAS.295..799W,1999A&A...352L..11H} but argued
to have an optical-V band magnitude of 18 \citep[but see \citep{1990A&AS...83..459T}]{2018arXiv181000566V}.

\subsection{Polarization}
Like the other domains of observation, in polarization too, OJ 287 exhibits
frequent and high variation in the degree of linear polarization (PD) as well as
electric vector polarization angle (EVPA) at radio \citep[e.g][]{2011ApJ...726L..13A,
2018ApJ...862....1C} and optical energies \citep{2000A&AS..146..141P,2016ApJ...819L..37V,
2017MNRAS.465.4423G,2019AJ....157...95G}. The reported PD range spans $\sim$ 0-40\%
at optical, $\sim$ 1.5-20\% at IR \citep{1984MNRAS.211..497H,2011ApJ...726L..13A,
2017MNRAS.465.4423G} and $\sim$ 0-30\% at radio \citep[e.g.][]{2011ApJ...726L..13A}
while PA changes of up to $\sim$ 250$^\circ$ has been observed \citep[e.g.][]
{2000A&AS..146..141P,2017MNRAS.465.4423G}. 

A detailed study by \citet{2010MNRAS.402.2087V} has reported
a preferred direction for optical EVPA which is consistent with the radio
measurements. They successfully decomposed the EVPA variations and found that
the preferred EVPA is due to the optically polarized core which is variable on
timescales of a year with a chaotic jet component superimposed on it. Similar PA
in radio and optical has also been reported by \citet{2018ApJ...864...67S} from
the analysis of quasi-simultaneous radio and optical observations. \citet{1984MNRAS.211..497H}
have reported a complex variation of polarization showing variation with
time as well as frequency (energy). Contrarily, close synchronous change of optical
and radio polarization has also been observed \citep[e.g.][]{1988A&A...190L...8K,
2009ApJ...697..985D}. Though generally optical PD follows the source
flux state, systematic to chaotic and then back to systematic changes in fractional
Stoke polarization during high flux states has also been observed \citep{2019AJ....157...95G}.
Study of EVPA time series by \citet{2018ApJ...862....1C} over 40 years,
assuming smooth changes, have found four rotation reversals in EVPA: $\sim$
180$^\circ$ anti-clockwise swing followed by a clockwise swing by the same
amount. The swings have taken place over timescales of a few weeks to a year.

\section{Discussion}\label{sec:discussion}
The observational results listed above clearly demonstrate that OJ 287 is a very
dynamic BLL object with activity over vast scales and across the observational domains.
Except for a few rare cases, these results, in general, appear to be random events
without any connections, reflective of stochastic variations of blazars as well
as the lack of a comprehensive jet theory. Additionally, it also highlights the
complexity of exploring non-linear dynamics involving a multitude
of scales both theoretically and observationally \citep[e.g.][]{2016ARA&A..54..725M}.
The best example of this complexity in the present case can be understood from studies
of source with VLBA. \citet{2012ApJ...747...63A} study of 1995-2011 radio images at
43 GHz argue for an erratic jet variation. On the contrary, an analysis of 15 GHz VLBA
images between 1995--2017 by \citet{2018MNRAS.478.3199B} have reported a periodicity of
$\sim$ 22-yr assuming that observed changes are smooth while \citet{2017Galax...5...12C}
have argued a periodicity of $\sim$ 30-yr using the same 15 GHz images but between
1995--2015 and employing ridge-line contours as an observational indicator.
Thus, given our current understanding of AGNs constituents,
their energy-dependent emission, and observational indications of a multitude of scales,
both long and short term monitoring across the EM bands is essential to unravel
the various facets and the source as a whole.

As stated already, the \(\sim 12\)-yr QPOOs has been one of the most explored
features of the source. Regarding the two class of models/interpretations for
the $\sim$ 12-yr QPOOs, the model of \citet{1996ApJ...460..207L}
which attributes the outbursts to thermal bremsstrahlung emission as a result of
disk-impact, is dynamical and predictive (the timing of the outbursts). Thus,
the predicted timing of the outbursts, separation between the two outbursts, optical
PD (unpolarized), and outbursts' temporal profile can be compared with observations
\citep[e.g.][]{2016MNRAS.457.1145P,2016ApJ...819L..37V,2018ApJ...866...11D}. Though
it predicts the outbursts to be unpolarized, given the
underlying jet emission, PD is expected to be systematically low compared to its
state before the outburst. From the timing of these outbursts \citep{2013A&A...559A..20H},
the model derived SMBH masses are $1.83\times10^{10}~M_\odot$ and $1.5\times10^{8}~
M_\odot$ for primary and secondary respectively \citep{2012MNRAS.427...77V,
2016ApJ...819L..37V,2018ApJ...866...11D}. The extreme mass of the primary, inconsistent
with estimates from other methods \citep[e.g.][and references therein]{2018MNRAS.478.3199B}
is, in fact, the biggest criticism of this model. The alternative interpretation
i.e. the geometrical class of models attributes the QPOs to Doppler boosted jet
synchrotron emission resulting from the jet precession \citep[e.g.][]{1997ApJ...478..527K,
1998MNRAS.293L..13V,2007Ap&SS.310...59S,2018MNRAS.478.3199B,2018arXiv181111514Q}.
Further, geometrical models are kinematic
in the sense that they are mainly concerned with reproducing the
QPOs and lack any predictive power of the timing of these outbursts. These models
mainly argue the central engine mass in the range of a few $\rm \times(10^{7}-
10^{9})~M_\odot$ \citep{2007Ap&SS.310...59S,2018MNRAS.478.3199B,2018arXiv181111514Q}.
However, it should be noted the system mass in these models is not dynamically
connected with the model parameters and have been mainly inferred or argued using other 
observations \citep[and references therein]{2007Ap&SS.310...59S,2018MNRAS.478.3199B}.
Thus, if the geometrical interpretation is true, the outbursts should be, in general,
highly polarized. Also, there should be a coherent simultaneous rise of emission
across the EM spectrum with similar temporal profiles. An important caveat in this
interpretation, however,
is that even jet emission has been observed to show low PD during flares and thus,
need statistically relevant polarization data to test either class of models.
Additionally, OJ 287 being a BLL object, knowing the expected time of outburst allow a targeted
observation campaign to study HE emission mechanisms -- one of the focus of intense
research in the blazar community in the \emph{Fermi} era. In the context of these two, the
MW observation of the 2015 outburst provides new, independent clues as argued below.

As per the prediction for the 2015 outburst in the disk-impact SBBH model \citep[November
2015 -- January 2016,][]{2010ApJ...709..725V}, an outburst with the expected signature --
flux peak coincident with a dip in the PD \citep[$<$ 10\%;][]{2016ApJ...819L..37V,
2017MNRAS.465.4423G} and broadly following the expected temporal profile of a
sharp rise followed by a slower decline with multiple smaller outbursts, was observed on
December 5th, 2015 \citep[MJD: 57361;][]{2016ApJ...819L..37V,2008Natur.452..851V}.
Additionally, a coincident large systematic swing of
$\sim$ 200$^\circ$ in the optical EVPA was also observed, similar to the EVPA swing
during the 1994 outburst \citep{2000A&AS..146..141P}. The most important point about
2015 outburst, however, was that it was the first QPOO with a true broadband coverage
from radio to MeV-GeV energies, thanks to the \emph{Fermi}.
Previous observation by \emph{CGRO} of the 1994 outburst only claimed a probable
detection in a ``hard state'' without any spectral study \citep{1996A&AS..120C.599S}
while 2007 observations by the VHE facility \emph{MAGIC} resulted in only upper
limit \citep{2009PASJ...61.1011S}. A detailed systematic study of broadband
activity around the 2015 QPOO by \citet{2018MNRAS.473.1145K} reported rise/outburst
at X-ray and MeV-GeV energies as
well. The observed MW variations were typical of the source, showed simultaneous
variability from NIR to $\gamma$-ray energies \citep{2018MNRAS.473.1145K} but 
broadband SEDs revealed a new spectral state of the source characterized by a break in
the NIR-optical SED and a hardening of MeV-GeV emission with a shift in its peak
emission. Further spectrotemporal analyses revealed that the NIR-optical break first
appeared on 27 May 2013 (MJD 56439) within the available NIR-optical data and 
is fairly well constrained in the sense that records jut before this i.e. MJD
56434 and before do not show such spectral break. It should, however, be
noted that the data gaps before and after MJD 56439 do not allow to track the
spectral evolution, which on first sight appears as though the spectral break is
due to drop of the two data points (J \& K bands; ref fig. \ref{fig:msSEDs}-c).
In fact, a power-law spectrum from these two data points already hints a marginal
excess even before MJD 56439 but is unreliable given that low-energy hump peaks
near these energies which may wash out the intrinsic spectrum due to smoothing
by change in spectral shape.

As for spectral changes and HE emission is concerned, multiple explanations have
been proposed. \citet{2018MNRAS.473.1145K} showed that an EC-BLR reproduces the
MeV-GeV emission while NIR-optical break nicely fits the description of a standard
disk emission of a $\sim 10^{10}~M_\odot$ SMBH (ref \ref{fig:msSEDs}-(b)).
Subsequently, \citet{2019arXiv190403357Q} argued that NIR-optical break is
actually a shift in synchrotron peak and is in tune with the shift observed in HE SED
peak while \citet{2019ICRC...36..971O} has proposed a hadronic scenario for the
MeV-GeV. However, as shown in \citet{2019BHCB}, the SED corresponding to the impact
flare has similar NIR and X-ray emission and lacks any spectral change vis-a-vis
2009 jet SEDs \citep{2013MNRAS.433.2380K}. This 
rules out \citet{2019arXiv190403357Q} interpretation and also the previous EC-IR
explanation for the MeV--GeV emission \citep{2019BHCB}. The hadronic model, on the
other hand, mainly focuses on the interpretation of the MeV-GeV emission and the NIR-optical
break remains unexplained. Neither the optically thin bremsstrahlung emission
from a 25eV thermal plasma responsible for QPOOs, as argued in \citet{2019ApJ...882...88V}
is consistent with the NIR-optical break though it may have a sub-dominant contribution.
Optically thin bremsstrahlung emissivity has a spectral index of \(\sim 0\)
(\(F_\nu \sim \nu^{-a}\)). Thus, in the blazar SED representation (\(\nu F_\nu\))
it will have a spectral index of \(\sim -1\), contrary (rising) to the observed
optical-UV (declining) SED \citep[see][]{2012MNRAS.427...77V}. At most, the maximum
possible contribution can be the lowest flux value observed in the NIR-UV
bands i.e. UVOT-w2 band ($1.5\times10^{15}$ Hz). With this, the flux contribution
at NIR-optical junction ($\sim 10^{14}$ Hz) will be an order of magnitude below the 
UVOT-w2 band, contrary to the observed SED (see \ref{fig:msSEDs}-(b)). Further,
for IC, bremsstrahlung (and accretion-disk) photons energy density will appear
de-boosted by a factor of the square of the bulk Lorentz factor (\(\Gamma\)) of
the emission region. Also, the IC spectrum peak (\(\nu_p\)) will be at \(\nu_p =
\delta/(1+z)\gamma_b^2(\nu^*/\Gamma)\) where \(\delta\) is the Doppler factor,
\(\gamma_b\) is the Lorentz factor corresponding to the break in a broken
power-law particle distribution normally assumed to model blazars SEDs,
z is the source redshift, and \(\nu^*\) is the frequency at which the seed
photon spectrum peaks \citep[e.g.][]{2013MNRAS.433.2380K}. Using \(\nu^* 
\sim 7.2\times 10^{15} \) Hz from \citet[eq 68;][]{2016MNRAS.457.1145P} results
\(\nu_p \sim 2.2\times10^{22}\) Hz assuming \(\delta = \Gamma\) and \(\gamma_b = 2000\)
\citep{2018MNRAS.473.1145K,2013MNRAS.433.2380K} while the observed HE SED peak
is at \(\sim 10^{24}\) Hz \citep{2018MNRAS.473.1145K}. An additional clue against
a dominant contribution of bremsstrahlung photons in IC is that the MeV-GeV
spectral profile remains as it was during the 2015 QPOO \citep{2018MNRAS.479.1672K} 
even after the disappearance of the NIR-optical spectral break around March 3, 2016
(MJD 57455; fig. \ref{fig:msSEDs}-(d)).

Among the discussed HE emission mechanisms,
the EC-BLR explanation seems the best description in the view of the observational
records in the literature during the previous QPOOs. The interpretation is consistent
with the detection of broad emission lines
during the previous cycles (1984, 2005-2008) of the $\sim$ 12-yr optical outbursts
as well as the strong changes observed in its level of emission \citep{2010A&A...516A..60N}.
Thus, if we combine the current EC-BLR origin of MeV-GeV emission \citep{2018MNRAS.473.1145K}
with the previous EC-IR \citep{2013MNRAS.433.2380K} and the inability of SSC to
reproduce the X-ray and $\gamma$-ray emission \citep{2013MNRAS.433.2380K,2009PASJ...61.1011S},
these results imply that in OJ 287 the MeV-GeV emission is due to EC, both on long and short
timescales. Additionally, this also provides the first clear observational evidence on
the ongoing debate of the location of the blazar zone at sub-parsec and parsec scales.
Further, if we extrapolate current inferences of OJ 287 HE emission with inferences
from the modeling of neutrino event SEDs of blazar TXS 0506+056 \citep{2018NatAs.tmp..154G},
it seems that MeV-GeV emission in all blazars is likely IC in origin i.e. leptonic in
origin. 

The 2015 QPOO was followed by yet another new and peculiar MW activity,
as reported in detail by \citet{2018MNRAS.479.1672K}. It started almost immediately
after the settling of the activity associated with
the December 2015 outburst. The source was found to be in a historic high state in
X-rays and was concurrently detected at VHEs \citep{2017arXiv170802160O}. The systematic
study of broadband SEDs presented by \citet{2018MNRAS.479.1672K} established this
to the presence of an additional HBL emission component with the low-energy component
peak in UV-soft-X-ray region. They further showed that the broadband SEDs are broadly
consistent with a two-zone leptonic model with one emitting the typical OJ 287 emission
with the modified MeV-GeV
spectrum and the other an HBL spectrum. An important aspect during this activity was the
further hardening of the MeV-GeV gamma-ray spectrum compared to the previous December 2015
activity and also a change of fraction polarization from systematic to chaotic and
then back to systematic trends \citep{2019AJ....157...95G}.

The interpretation of the NIR-spectral break with an accretion disk emission has direct
implications on the ongoing debate over the central engine of OJ 287. The consistency
of the spectral break with an accretion disk spectrum of a $\sim 10^{10}~M_\odot$
SMBH provides independent evidence from the energy-spectrum domain in favor of the
disk impact SBBH model which is based solely on the QPOOs timing \citep{2016ApJ...819L..37V}.
Additional strong evidence in its favor is the close coincidence between the time
of appearance of the spectral break in May 2013 (MJD 56439) and the impact time predicted
by the model in the SMBH frame \citep{2018ApJ...866...11D}.  This is in stark
contrast with the geometrical class of models which argue for a total system-mass
in the range of $\rm \times(10^{7} - 10^{9}) ~M_\odot$. However, the geometrical
models are still plausible as the system-mass is not dynamically related to the
model parameters. Thus, precessing jet models with a total central-engine mass of
$\sim 10^{10}~M_\odot$ are still consistent with the currently available data/results
\citep{2018MNRAS.473.1145K} in the view of the lack of statistically relevant polarization
information. The biggest challenge with geometrical models, however, is their failure
to explain the sharpness of the outbursts \citep{2008Natur.452..851V}. Thus, if the
geometrical interpretation represents the real physics behind the phenomena, the
sharpness of profile during the recurring outbursts would indicate some special
physical processes happening during these repeating outbursts \citep[e.g.][]{2018arXiv181111514Q}.
Another result contrary to the geometrical class of models is the lack of similar
temporal profile at radio and optical \citep{2018MNRAS.478.3199B}.

The peculiarity and uniqueness of the $\sim 12$-yr QPOOs are further supported
by the X-ray records in the literature during and around these outbursts. Though
the concurrent NIR-optical observations during the previous cycles of $\sim$
12-yr QPOOs do not show any spectral break \citep{1971NPhS..234...71D,
1986Natur.324..546G,2000A&AS..146..141P}, an extremely soft X-ray spectral state
seems to be a generic feature within the limits of observational records.
It seems to be present within a few years around these optical outbursts
and shows strong variations in spectral extent, strength, and the time of appearance
\citep[and references therein]{2009PASJ...61.1011S,1997PASJ...49..631I}, thereby
suggesting a relation between the two. The 2016--2017 MW activity showing an additional
transient HBL component appears to be the continuity of this trend. The peculiarity
of these QPOOs is also noted from time series analysis  \citep{2018MNRAS.478.3199B}. Finally, it
should be noted that all the models are primarily based on the observed
periodicity and none currently reproduce all the observational features seen
during these outbursts \citep[e.g.][]{1996A&A...315L..13S}. The disk-impact
model lacks proper accounting of magnetic field effects which is essential for polarization properties.
Though the phenomenological interpretation for the observed PD during the 2015
outburst in \citet{2016ApJ...819L..37V} seems justified, it lacks the explanation
of the large systematic EVPA swing. 

\section{Summary}\label{sec:summary}
The new spectral features seen during the 2015--2017 MW high activity of OJ 287
have provided some tantalizing, independent clues in settling two of the active
ongoing debates on the source related to its central engine and MeV -- GeV emission
mechanism. Though different interpretations may be possible, the spectral coincidence
of the NIR-optical spectral break with a standard accretion-disk emission of a $\sim
10^{10}~M_\odot$ SMBH and its first appearance in close coincidence with the impact
time predicted by the disk-impact binary SMBH model in SMBH frame support/favor
the disk-impact binary SMBH model over the geometrical class of models. However,
geometrical models still cannot be ruled out confidently as the mass of the central
engine is not dynamically tied with the model parameters and hence, are still plausible
within the limits of currently available observational data. In this case, the sharpness
of QPOOs, as well as non-similarity of optical and radio time series need an explanation.
Future observations of these outbursts, particularly the duration between the twin outbursts
and polarization hold the key to break this degeneracy.

Similarly, the change of the MeV-GeV spectrum is reproduced by both
hadronic and leptonic scenarios. The leptonic scenario via EC-BLR, however, seems natural,
consistent with the detection of broad emission lines during the previous impact
duration and strong changes in its luminosity \citep{2010A&A...516A..60N}. This with the
inferences from the systematic broadband SEDs modeling during its previous activity
\citep{2013MNRAS.433.2380K} implies that
in OJ 287 the MeV-GeV emission is due to EC, both on long and short timescales. These
results also provide the first clear evidence on the debate of the location of the blazar
emission region at parsec and sub-parsec scales. The extremely soft X-ray spectral states
around the QPOOs make it an ideal target for the Cherenkov Telescope Array (CTA) -- the next
generation, ground-based gamma-ray observatory.

\vspace{6pt} 

\funding{The Author acknowledges funding from FAPESP grant number 2015/13933-0.}

\acknowledgments{ The author thanks the anonymous referees for their thorough reports
which helped in the improvement of the overall presentation.}

\conflictsofinterest{The authors declare no conflict of interest.} 

\abbreviations{The following abbreviations are used in this manuscript:\\

\noindent 
\begin{tabular}{@{}ll}
AGNs & Active Galactic Nuclei \\
BLL & BL Lacartae \\
BLR & Broad Line Region \\
CTA & Cherenkov Telescope Array \\
EVPA & Electric vector polarization angle \\
FSRQ & Flat Spectrum Radio Quasar \\
HSP (HBL) & High-synchrotron peaked (High-frequency peaked BL Lac) \\
IC & Inverse Compton \\
ISP (IBL) & Intermediate-synchrotron peaked (Intermediate-frequency peaked BL Lac) \\
LSP (LBL) & Low-synchrotron peaked (Low-frequency peaked BL Lac) \\
MW & Multi-wavelength \\
NIR & Near-infrared \\
PD & Polarization Degree \\
QPO & Quasi-periodic Outburst \\
QPOO & Quasi-periodic Optical Outburst \\
SED & Spectral Energy Distribution \\
SMBH & Suppermassive Black Hole \\
VHE & Very High Energy (E > 100 GeV)
\end{tabular}}


\reftitle{References}

\bibliography{pankaj_nthU_rv3.bib}
\externalbibliography{yes}

\end{document}